\documentclass[10pt,letterpaper]{article}
\usepackage[top=0.85in,left=2.75in,footskip=0.75in]{geometry}

\usepackage{changepage,color}  

\usepackage[utf8]{inputenc}

\usepackage{textcomp,marvosym}

\usepackage{fixltx2e}

\usepackage{amsmath,amssymb}

\usepackage{cite}

\usepackage{nameref,hyperref}

\usepackage[right]{lineno}

\usepackage{microtype}
\DisableLigatures[f]{encoding = *, family = * }  

\usepackage{rotating}

\usepackage{soul}
\usepackage{comment}

\raggedright
\usepackage[aboveskip=1pt,labelfont=bf,labelsep=period,justification=raggedright,singlelinecheck=off]{caption}

\makeatletter
\renewcommand{\@biblabel}[1]{\quad#1.}
\makeatother

\date{}

\usepackage{lastpage,fancyhdr,graphicx}
\usepackage{epstopdf}
\fancyheadoffset[L]{2.25in}
\fancyfootoffset[L]{2.25in}

\begin{document}
\vspace*{0.35in}

\begin{flushleft}
{\Large
\textbf\newline{Large scale chromosome folding is stable against local changes in chromatin structure}
}
\newline
\\
Ana-Maria Florescu\textsuperscript{1,\ddag,*},
Pierre Therizols\textsuperscript{2,3,4},
Angelo Rosa\textsuperscript{1,\ddag,**}
\\
\bigskip
{\bf{1}} SISSA - Scuola Internazionale Superiore di Studi Avanzati, Via Bonomea 265, 34136 Trieste (Italy)

{\bf{2}} INSERM UMR 944, {\'E}quipe Biologie et Dynamique des Chromosomes, Institut Universitaire d'H{\'e}matologie,   
H\^opital St. Louis, 1   Avenue   Claude Vellefaux, 75010 Paris (France) 
\\
{\bf{3}} CNRS UMR 7212, 75010 Paris (France)
\\
{\bf{4}} Universit\'e Paris Diderot, Sorbonne Paris Cit\'e, 75010 Paris (France)  
\bigskip

%
%
\ddag These authors contributed equally to this work.




* aflorescu@sissa.it,
** anrosa@sissa.it

\newpage
\end{flushleft}

\section*{Abstract}
Characterizing the link between small-scale chromatin structure and large-scale chromosome folding during interphase is a prerequisite for understanding transcription.
Yet, this link remains poorly investigated.
Here, we introduce a simple biophysical model where
interphase chromosomes are described in terms of the folding of chromatin sequences composed of alternating blocks of fibers with different thicknesses and flexibilities,
and we use it to study the influence of sequence disorder on chromosome behaviors in space and time.
By employing extensive computer simulations,we thus demonstrate that chromosomes undergo
noticeable conformational changes only on length-scales smaller than $10^5$ basepairs and time-scales shorter than a few seconds,
and we suggest there might exist effective upper bounds to the detection of chromosome reorganization in eukaryotes.
We prove the relevance of our framework by modeling recent experimental FISH data on murine chromosomes.


\section*{Author Summary}
A key determining factor in many important cellular processes as DNA transcription, for instance,
the specific composition of the chromatin fiber sequence has a major influence on chromosome folding during interphase.
Yet, how this is achieved in detail remains largely elusive.
In this work, we explore this link by means of a novel quantitative computational polymer model for 
interphase chromosomes where the associated chromatin filaments are composed of mixtures of fibers with heterogeneous physical properties.
Our work suggests a scenario where chromosomes undergo only limited reorganization,
namely on length-scales below $10^5$ basepairs and time-scales shorter than a few seconds.
Our conclusions are supported by recent FISH data on murine chromosomes.


\section*{Introduction}

Understanding how genomes fold within the crowded environment of the nucleus~\cite{Alberts} during interphase
represents a necessary step for the comprehension of important cellular processes such as gene expression and regulation~\cite{BickmoreReview2015}.
The combined results of high-resolution microscopy~\cite{SuperResGeneralReview,RicciCosmaCell2015} and mathematical and computer modelling~\cite{MartiRenomMirnyRev2011,RosaZimmer2014}
seem to suggest that genomes are organized hierarchically~\cite{WoodcockDimitrov2001,BancaudLavelle2012}.
Each genome is partitioned into a set of single units, the chromosomes, and each chromosome is made of a single filament of DNA complexed
around histone octamers to form a necklace-like fiber $\approx 10$ nm thick known as the 10nm chromatin fiber.
In {\it in vitro} conditions close to the physiological ones, this fiber is observed to fold into a thicker, more compact structure known as the 30nm fiber~\cite{Alberts},
whose role and existence {\it in vivo} are nonetheless still quite debated~\cite{ReviewChromosoma2015,MaeshimaEltsov}.
On larger scales,
chromosome conformation capture (3C) techniques~\cite{BickmoreReview2015} have shown that chromosomes appear organized in Topologically Associated Domains (TADs)
of sizes ranging from $\approx 0.1$ to $\approx 1$ megabasepairs (Mbp).
Chromosome loci within TADs interact frequently between themselves, but less frequently across different TADs.
Finally, chromosomes do not spread inside the whole nucleus, rather they occupy well localized nuclear regions (the so-called ``chromosome territories'') which play a crucial role in gene expression and regulation~\cite{CremerReview2001}.

While much progress on the causal relationship between chromosome structure and function has now been made,
many fundamental aspects remain still obscure.
One of these crucial issues concerns the link between chromosome (re)organization at various length and time-scales and
the spread of inhomogeneties in the sequence of the chromatin fiber which may arise from
(1) selective epigenetic marks induced by chemical modifications of the histone tails~\cite{BoettigerZhuang2016},
(2) nucleosomes arrangements in discrete nanodomains of different sizes \cite{RicciCosmaCell2015},
and
(3) selective stimulation of particular kinds of genes~\cite{TherizolsBickmoreScience2014} or entire gene families~\cite{KocanovaBystricky2010}.
These events will produce modifications in the local polymer properties of the chromatin fiber (as its persistence length or the local compaction ratio) which might affect in turn the whole hierarchical folding of chromosome organization~\cite{ReviewChromosoma2015,MaeshimaEltsov}.

Motivated by these considerations,
we present here the results of Molecular Dynamics computer simulations of a minimal polymer model for interphase chromosomes in order to quantify to what extent
the simultaneous presence of chromatin fibers of heterogeneous composition (different thicknesses and flexibilities)
is able to generate observable effects on the small- and large-scale structures and motions of the associated chromosome.

The proposed model complements previous work by one of the authors~\cite{RosaPLOS2008,RosaBJ2010,DiStefanoRosa2013,ValetRosa2014} concerning the description of chromosome folding in terms of fundamental polymer physics.
Similarly to other recent works discussing the explicit role of sequence disorder in chromatin~\cite{Diesinger2010} and chromosome behaviors~\cite{NicodemiPombo2012,JostVaillantNAR2014},
here we ``deviate'' from the description of the chromatin filament as a {\it homopolymer} and we discuss {\it sequence effects in space and time} through the introduction of controlled amounts of disorder in the chromatin sequence.
In this way, we provide a {\it quantitative} description for many crucial aspects concerning the structure and dynamics of interphase chromosomes
which are ``spontaneously'' driven by the physical properties of the underlying chromatin sequence with a definite {\it copolymer} structure.

In particular, by considering the two ``extreme'' cases of chromosomes made of:
(1)
short stretches of a thinner, more flexible fiber randomly interspersed in a ``sea'' of thicker fiber
and
(2)
chromosomes partitioned into two distinct blocks of thinner and thicker fibers,
we show that significative spatial and dynamical rearrangements of chromatin loci
appear to be restricted to limited contour lengths (up to $\approx 10^5$ basepairs (bp)) and times scales (up to few seconds).
Interestingly, there exists a limited range ($10^4-10^5$ bp) of chromatin contour lengths where chromosome expansion is accompanied by an increase (rather, than a decrease) of the associated contacts between the fibers.

We apply this framework to rationalize the outcome of recent experiments which employ fluorescent microscopy
to monitor conformational changes of chromosomes that occur upon transcription activation or chromatin decondensing in mouse embryonic stem cells~\cite{TherizolsBickmoreScience2014}.
Finally, we argue that the effects discussed here are not the consequence of the details of the model, but involve more general aspects of polymer physics.

\section*{Results}

In this section, we present the main results of our model by addressing in particular the specific question of how modifications in the small-scale properties of the chromatin fiber
turn to affect chromosome behaviour on much larger scales.
In order to model these modifications, we represented
the linear chromosome filament
as a {\it heterogeneous copolymer} made of two types of fibers with physical properties matching those of the 10nm and the 30nm chromatin fibers:
10nm fibers are modelled as completely flexible while 30nm fibers have a persistence length $=150$ nm as in previous works~\cite{RosaPLOS2008,RosaBJ2010,DiStefanoRosa2013,ValetRosa2014}.
Then, we considered several situations:
either short ($3000$ bp) fragments of 10nm fibers are distributed randomly along the whole chromatin filament,
or we studied chromosome structures where a single continuous 10nm filament is centered around the chromatin portion closer to or farther from the chromosome center of mass.
For more details and technical issues, we invite the reader to look into the ``Materials and Methods'' section.

With its true existence {\it in vivo} appearing more and more debatable~\cite{ReviewChromosoma2015,MaeshimaEltsov},
employment of 30nm fibers might look inappropriate.
However, our choice of physical parameters for the model was finally motivated by their previous use and proven consistency~\cite{RosaPLOS2008,RosaBJ2010,DiStefanoRosa2013,ValetRosa2014} with experimental data.
We posit though that our results should remain qualitatively valid for more general models of non-homogeneous polymers.

In order to quantify chromosome changes,
we have considered either spatial relationships between distal fragments along the chromatin sequence or the dynamic behaviour of chromatin loci.
The former aspects were investigated by focusing mainly on the following two observables:
(1)
the mean-squared internal distance ($\langle R^2(L) \rangle$) and
(2)
the average contact frequency $\langle p_c(L) \rangle$ between {\it all possible pairs} of genomic loci {\it at any given genomic separation $L$}.
For $\langle p_c(L) \rangle$, we have adopted the choice that two monomers are in contact whenever their spatial distance is smaller than a conventional cut-off distance $=60$ nm,  corresponding to twice the thickness of the 30nm fiber.
These two quantities are of particular experimental interest:
in fact, $\langle R^2(L) \rangle$ can be measured through fluorescence {\it in-situ} hybridization (FISH)~\cite{CremerReview2001},
while $\langle p_c(L) \rangle$ is the result of chromosome conformation capture (3C) techniques~\cite{dekker3c,DekkerMartiRenomMirny2013}.
Moreover, $\langle R^2(L) \rangle$ and $\langle p_c(L) \rangle$
are useful to distinguish between complementary aspects of chromosome conformation, as was recently highlighted by Williamson {\it et al.}~\cite{WilliamsonBickmore2014}.
It is then interesting to monitor our systems by employing both tools.
Dynamical aspects were instead discussed in terms of the mean-square displacement $\langle \delta r^2(\tau) \rangle$ of chromatin loci at lag time $\tau$.
This is also a quantity of notable experimental interest, as specific chromatin loci can now be followed {\it in vivo} by, {\it e.g.}, fluorescent microscopy on GFP tagged chromosome sequences~\cite{Gasser2001,MeisterGasserRev2010}.

All results are presented as functions of the basepair content of the 10nm fibers present in the model chromosomes {\it vs.} the total chromosome basepair content, expressed in {\it percent}
(see section ``Materials and Methods'' for details).
For comparison, model chromosomes with no ($0\%$) or entirely made of ($100\%$) 10nm fibers are also discussed.

\paragraph*{Large scale chromosome behaviour does not depend on the fiber composition.}

\begin{figure}[h]
\includegraphics[width=\textwidth]{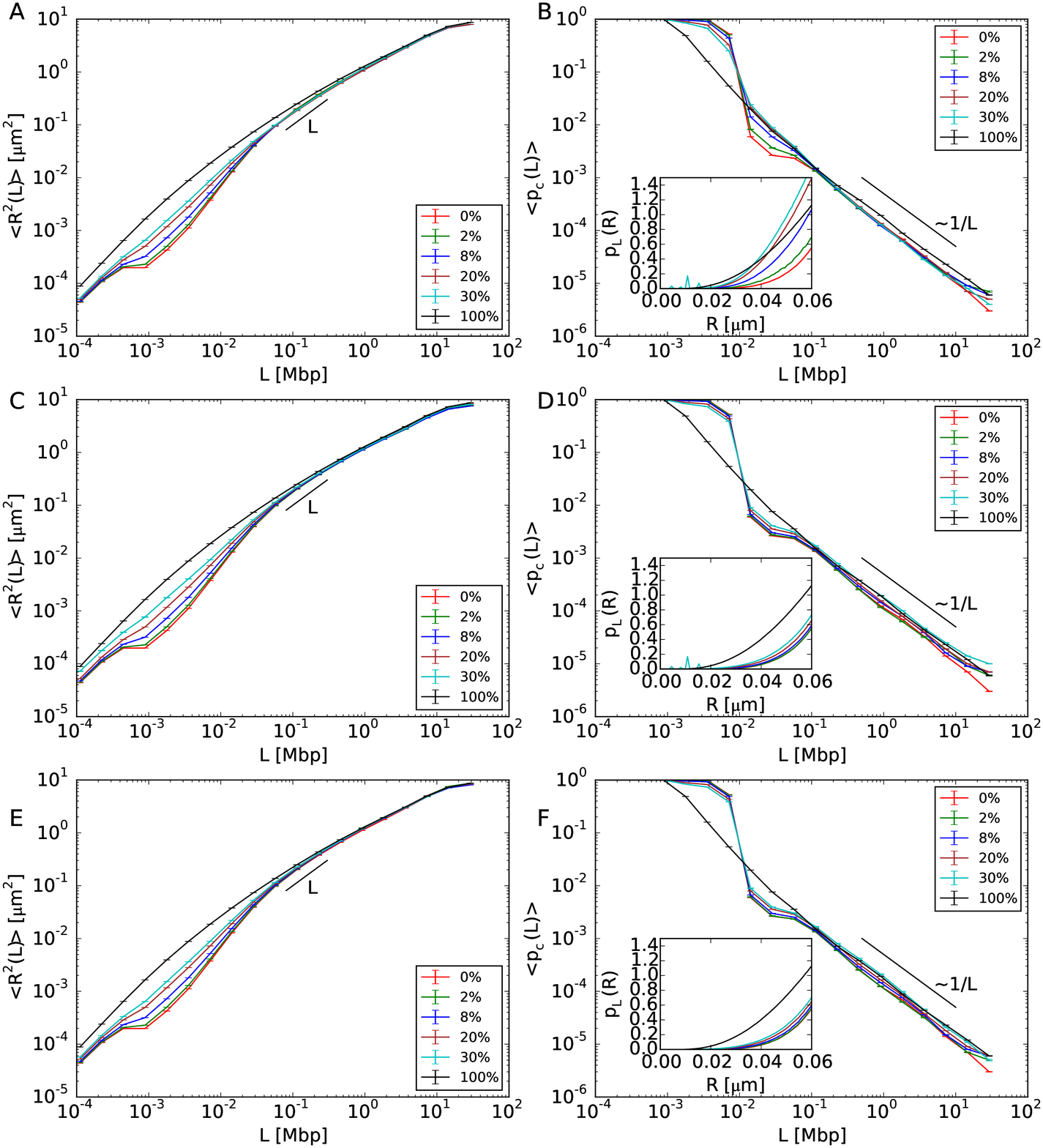}
\caption{{\bf Structural properties of model chromosomes composed of both 10nm fiber and 30nm fiber.}
The legends report the total amounts of 10nm fiber present in the corresponding model chromosomes.
(A, C, E)
Mean-square spatial distances, $\langle R^2(L) \rangle$, between pairs of chromosome loci separated by the genomic distance $L$.
(B, D, F)
Corresponding average contact frequencies, $\langle p_c(L) \rangle$.
The different panels correspond to the different case studies of random locations of 10nm chromatin fibers (top), and the continuous filament of 10nm chromatin fiber located closer (middle) and farther (bottom) from the chromosome center of mass.
Insets of panels B, D, F:
Small-$R$ behavior (up to the cut-off distance of $60$ nm) of the probability distribution function $p_L(R)$ of spatial distances $R$ at genomic separation $L=0.015$ Mbp,
which governs the behavior of $\langle p_c(L=0.015 \mbox{ Mbp}) \rangle$ (see equation~\ref{eq:PcEnd2EndPDF}).
}
\label{fig:rsqcprob}
\end{figure}

The results of our three case studies are summarized in figure \ref{fig:rsqcprob}.

It is visible that only length-scales smaller than $L \approx 0.1$ megabasepairs (Mbp) are affected with $\langle R^2(L) \rangle$ expanding sensibly more than in the situation where chromosomes are composed only of 30nm fiber (panels A, C, E),
while the behaviour at large scales remains unaffected.
Moreover, in the case where the unfolded sequences are grouped into a single cluster (middle and bottom panels),
there is no dependence on their spatial positioning with respect to the center of mass of the corresponding chromosome territory.

Insensitivity of large scales to changes at small ones is also confirmed (panels B, D, F) by the analysis of contact frequencies, $\langle p_c(L) \rangle$,
whose trend remains, in particular, compatible with the experimentally observed power-law $\langle p_c(L) \rangle \sim L^{-1}$~\cite{hic}.
Interestingly, instead of decreasing as expected from panels A, C and E,
corresponding contact frequencies in the limited range $[0.01 \mbox{ Mbp} - 0.1 \mbox{ Mbp}]$ also increase as a function of the 10nm fiber content (see in particular panel B).
In order to understand this result,
we use the mathematical relationship~\cite{RosaBJ2010} between average contact frequencies $\langle p_c(L) \rangle$ and distribution function $p_L(R)$ of internal distances $R(L)$ given by:
\begin{equation}\label{eq:PcEnd2EndPDF}
\langle p_c(L) \rangle = \frac{\int_0^{r_c} \, p_L(R) \, 4 \pi R^2 \, dR}{\int_0^\infty \, p_L(R) \, 4 \pi R^2 \, dR} \, ,
\end{equation}
where $r_c = 60$ nm is the cut-off distance for two monomers to form a contact.
In particular, equation~\ref{eq:PcEnd2EndPDF} implies that $p_L(R)$ should also increase as a function of the 10nm fiber content at given $L$.
The inset of panel B confirms this behavior for $L=0.015$ Mbp.
This specific trend is due to the decreasing of the chromatin ``effective persistence length'' (from $\approx 10^4$ to $\approx 10^2$ basepairs for model chromosomes entirely made of 30nm and 10nm fibers, respectively)
following from increasing amounts of 10nm fibers, which allows genomic loci to contact each other with enhanced probability.

For a more quantitative view on chromosome reorganization at small scales,
the same data for $\langle R^2(L) \rangle$ and $\langle p_c(L) \rangle$ were normalized to corresponding values for chromosome conformations made of 30nm fiber, see Figure 9.
We highlight in particular the pronounced peaks in the left panels corresponding to a maximum volume expansion of $\approx 30$,
and we pinpoint again the increasing of contact frequencies in the aforementioned interval $[0.01 \mbox{ Mbp} - 0.1 \mbox{ Mbp}]$.

The latter, in particular, constitutes a result with non trivial experimental implications.
First, in connection to some recently proposed protocols for the reconstruction of chromosome conformations based on 3C (reviewed in~\cite{Serra2015}) where a monotonous relationship between chromatin distances and contacts is often assumed,

our finding demonstrates that some caution is needed in order to avoid systematic bias in the final reconstructed structure.
Second, the recent puzzling result~\cite{WilliamsonBickmore2014} where chromatin domains with a high propensity to form 3C contacts seem to undergo rather pronounced decondensation when monitored by using FISH
can be understood by considering that the two experimental techniques sample very different intervals of the corresponding $p_L(R)$'s:
close to the average value or the median for FISH, around to the lower tail for 3C techniques, as shown in the insets of figure~\ref{fig:rsqcprob}B, D and F.

Our work thus supports the important conclusion of references~\cite{WilliamsonBickmore2014,MirnyFebs2015},
namely that one needs to take into account both kinds of data to reconstruct correctly the shape of chromatin domains.

\begin{figure}[h]
\includegraphics[width=\textwidth]{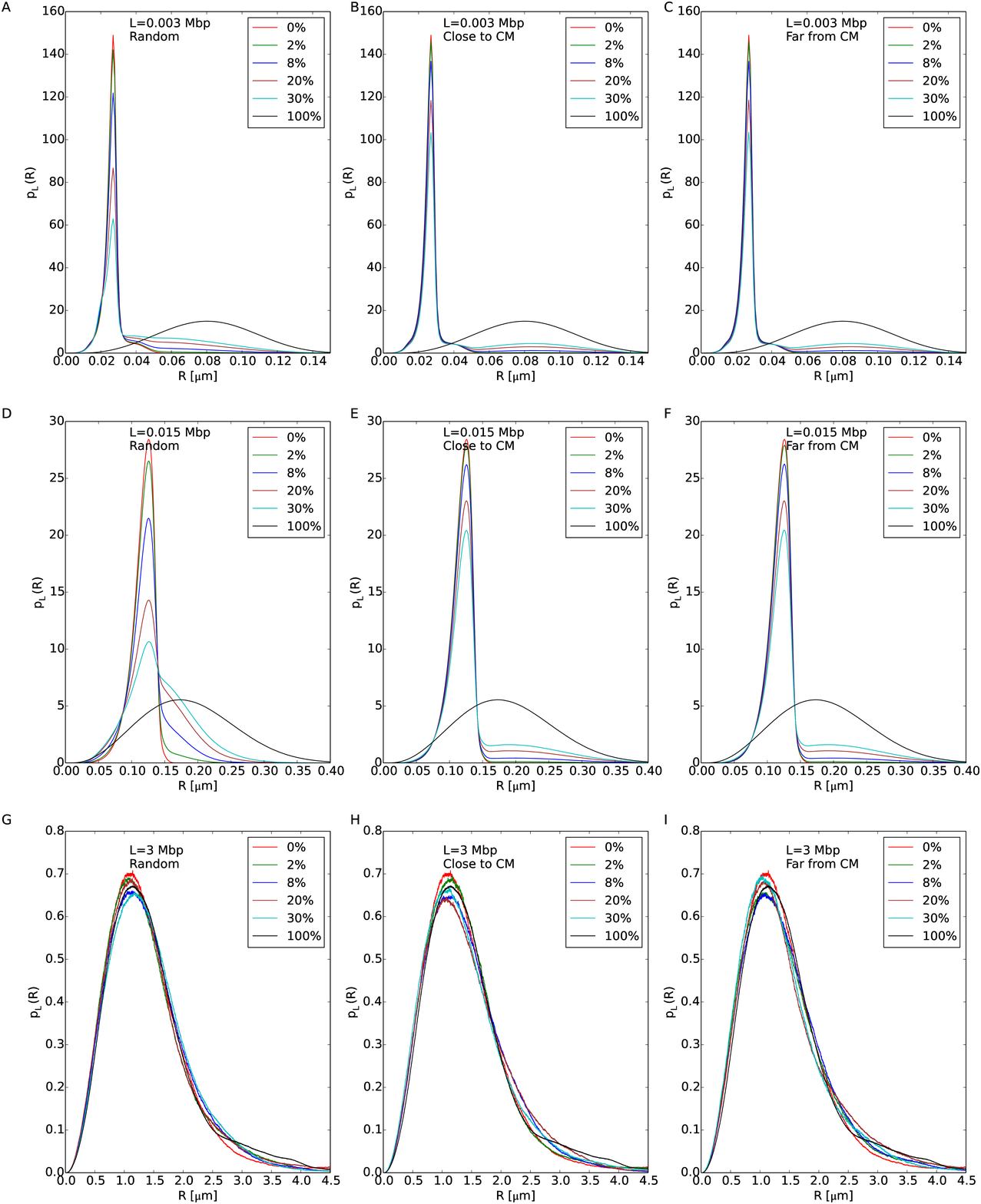}
\caption{{\bf Distribution functions of spatial distances $R$ at fixed genomic separations $L = 0.003, 0.015$ and $3$ Mbp (top, middle and bottom panels, respectively).}
The legends report the total amounts of 10nm fiber present in the corresponding model chromosomes.}

\label{rhist}
\end{figure}

We complete the discussion by considering the full distributions $p_L(R)$ for $L=0.003, 0.015$ and $3$ Mbp, see figure~\ref{rhist}.
For $L=0.003$ and $L=0.015$ Mbp (panels A-F) there are quantitative differences between the cases where the 10nm fiber is located at sparse random positions along the chromosome and where they form a single chromatin cluster,
while for $L=3$ Mbp (panels G-I) all distributions show no noticeable difference.
In panels A-F, the largest peak corresponds to the most probable value of spatial distances between loci on the 30nm fiber region.
Additionally, for random locations of 10nm fiber (panels A and D) corresponding $p_L(R)$'s show a broader population of $R$ values,
while in the other cases there exist smaller peaks corresponding to the most probable distance between loci on the 10nm fiber region.
Interestingly, similar distribution functions for spatial distances between chromatin loci seem to have been reported in yeast (see panels B, C, D, and E of figure 1 in reference~\cite{bystricky}).
Although a direct comparison between these experimental results and our data is not possible (our setup applies to large chromosomes, like mammalian ones),
we are tempted to speculate that the results reported in reference~\cite{bystricky} are a manifestation of the presence of chromatin fibers of different compositions.
Alternatively, we may interpret the observed shape of the $p_L(R)$'s in terms of a bias towards large spatial distances:
in that respect,
we report that a similar feature has been observed in human and mouse chromosomes~\cite{KocanovaBystricky2010,WilliamsonBickmore2014} (including the experimental data discussed in this work, see figure~\ref{fig:TherizolsData}, left panels).

\paragraph*{The mechanical behavior of the thinner fibers is influenced by the physical properties of the rest of the chromosome.}

\begin{figure}[h]
\includegraphics[width=1.1\textwidth]{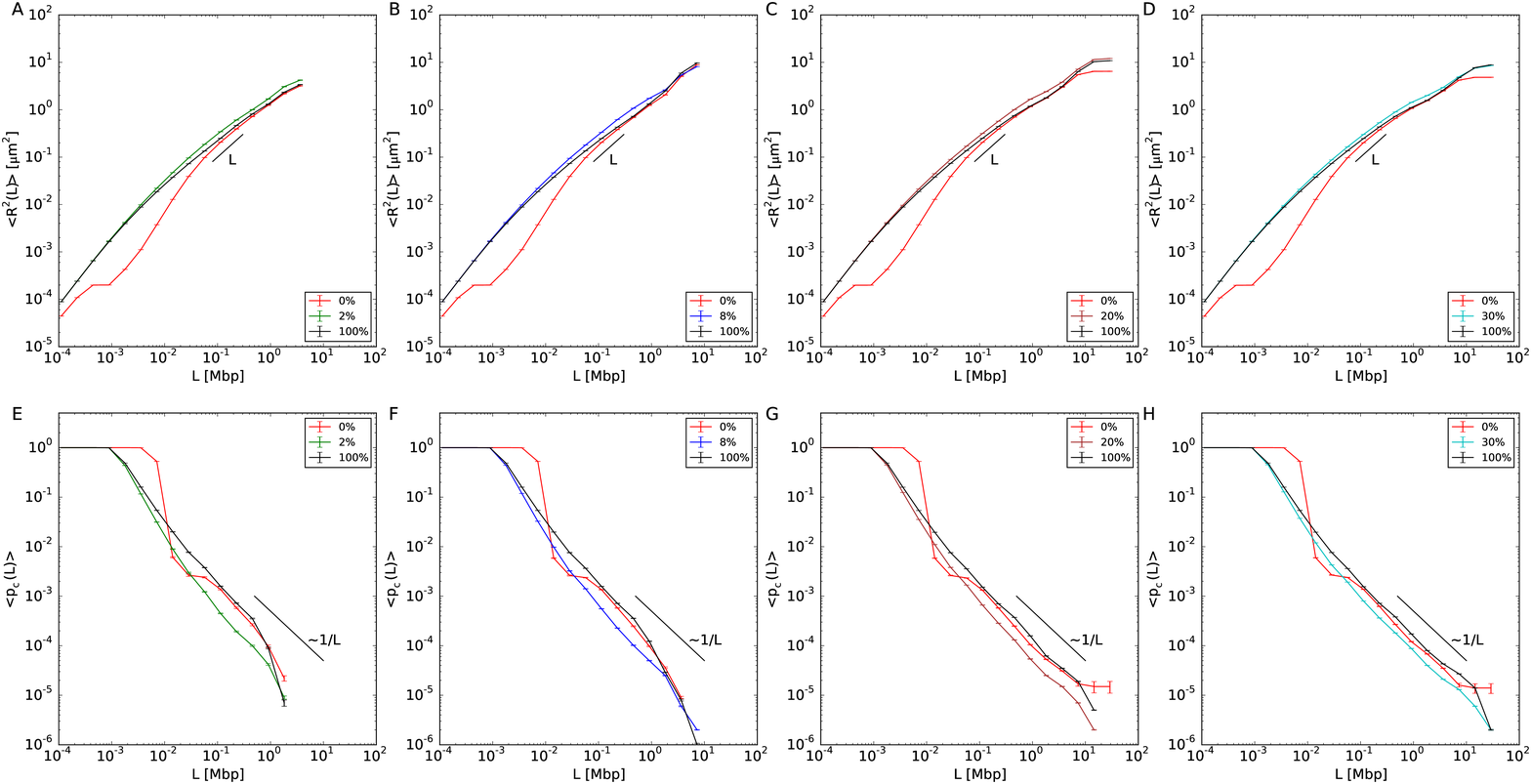}
\caption{{\bf Structural properties relative to the 10nm portion of model chromosomes composed of two separate domains of 10nm fiber and 30nm fiber, with the 10nm domain positioned closer to the chromosome center of mass.}
(Top panels) Mean-square spatial distances, $\langle R^2(L) \rangle$, between 10nm fiber loci separated by genomic distance $L$ for different model chromosomes (see legends) and in comparison to results obtained on the corresponding same sequences on model chromosomes with no ($0\%$) or entirely made of ($100\%$) 10nm chromatin fiber.
(Bottom panels) Corresponding mean contact frequencies, $\langle p_c(L) \rangle$.}
\label{fig:inside}
\end{figure}

\begin{figure}[h]
\includegraphics[width=1.1\textwidth]{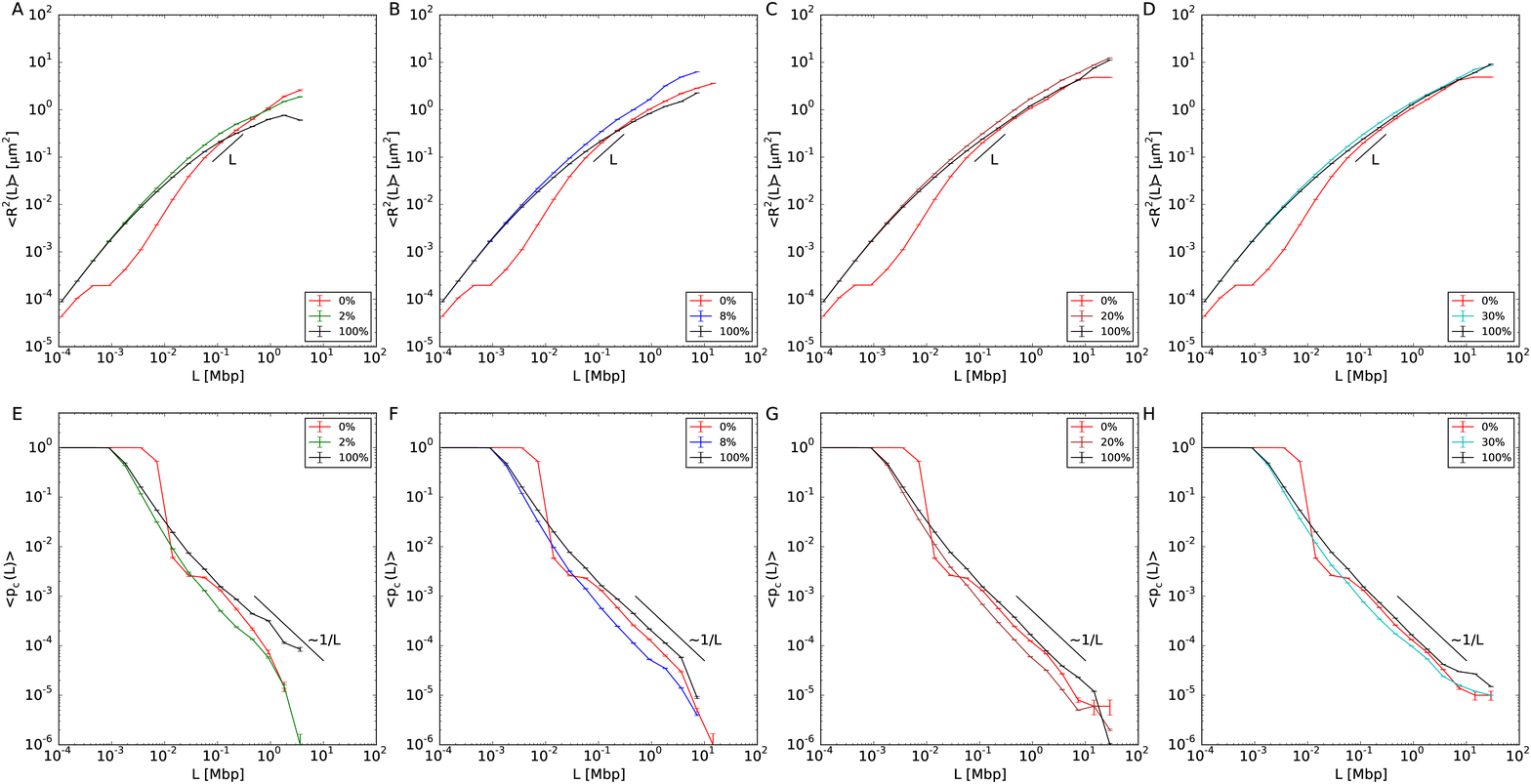}
\caption{{\bf Structural properties relative to the 10nm portion of model chromosomes composed of two separate domains of 10nm fiber and 30nm fiber, with the 10nm domain positioned farther from the chromosome center of mass.}
Symbols and color code are as in figure~\ref{fig:inside}.
}
\label{fig:outside}
\end{figure}

We have then considered those chromosome configurations with single long sequences of 10nm fiber
and we have calculated $\langle R^2(L) \rangle$ and $\langle p_c(L) \rangle$ on these sequences, see figures \ref{fig:inside} and \ref{fig:outside}.
Our results demonstrate that $\langle R^2(L) \rangle$ increases systematically with respect to the analogous quantity for model chromosomes made of 30nm fiber,
with no dependence on the specific location of 10nm fiber along the chromosome.
This is also shown by the decreased contact probability.
This insensitivity to positioning follows from the proposed picture~\cite{RosaEveraersPRL2014} that chromosomes resemble a uniformly dense, semi-dilute solution of branched polymer rings.
We expect then chromatin filaments to be similarly constrained or accessible regardless of their position along the genomic sequence.

In order to prove that the reported volume increase is not a fortuitous coincidence of the chosen sequence,
we have also repeated the analysis on the same genomic region for the model chromosome with no 10nm fiber ($0\%$) and the model chromosome entirely made of 10nm fiber ($100\%$).
Both $\langle R^2(L) \rangle$ and $\langle p_c(L) \rangle$ differ quantitatively from the case in which just one part of the chromosome is allowed to swell.
Similarly to the previous section, the same results can be recast in terms of ratios to the corresponding quantities calculated for chromosome conformations with no 10nm fiber,
see Figure 10 and Figure 11, which allows to appreciate to what extent chromosomes may effectively reorganize.

Because of the copolymer structure of our model chromosomes,
the two types of fibers have different stiffnesses and thicknesses which cause the thinner fiber to move away from the thicker one.
Consequently, $\langle p_c(L) \rangle$ calculated for the 10nm fiber sequence tends to decrease and shows a scaling behavior $\sim L^{-1.18 \pm 0.05}$ for $L$ in the interval $0.01 - 1$ Mbp
which, being slightly steeper than the one for the entire chromosome, suggests an increase in the volume spanned by the fiber.
Qualitatively, volume differences between chromosome regions with different transcription activities have been reported in a recent study on {\it Drosophila}~\cite{BoettigerZhuang2016}.
Here the authors have studied three regions of the {\it Drosophila} genome, which, according to the histone modifications that they were bearing, were classified as active, inactive and polycomb-repressed.
They found that the polycomb-repressed region was the most compact, followed by the inactive region, while the active region was the least compact.
These results encourage us to speculate that the difference between the active and inactive regions could be due to different polymer physical properties induced by the histone modifications.

\paragraph*{Single clusters of thinner fiber increase the mobility of corresponding genomic loci.}

Here we discuss the impact of chromatin unfolding on the dynamics of the corresponding genomic loci.
Specifically, we have considered the mean-square displacement $\langle \delta r^2(\tau) \rangle \equiv \langle ({\vec r}_i(t+\tau) - {\vec r}_i(t))^2 \rangle$ at lag time $\tau$,
where $\vec r_i(t)$ is the spatial position of monomer $i$ at time $t$ and we implicitly assume average over specific monomer positions along the chromatin chain.
In fact, this is an observable widely employed in many experiments monitoring the dynamic activity of specific chromatin loci,
being especially suitable for comparing genome behavior in response to changes of the environment~\cite{ZimmerNature2006},
or when the cell is targeted with drugs which are able to activate selectively certain types of genes~\cite{KocanovaBystricky2010}.

\begin{figure}[h]
\includegraphics[width=\textwidth]{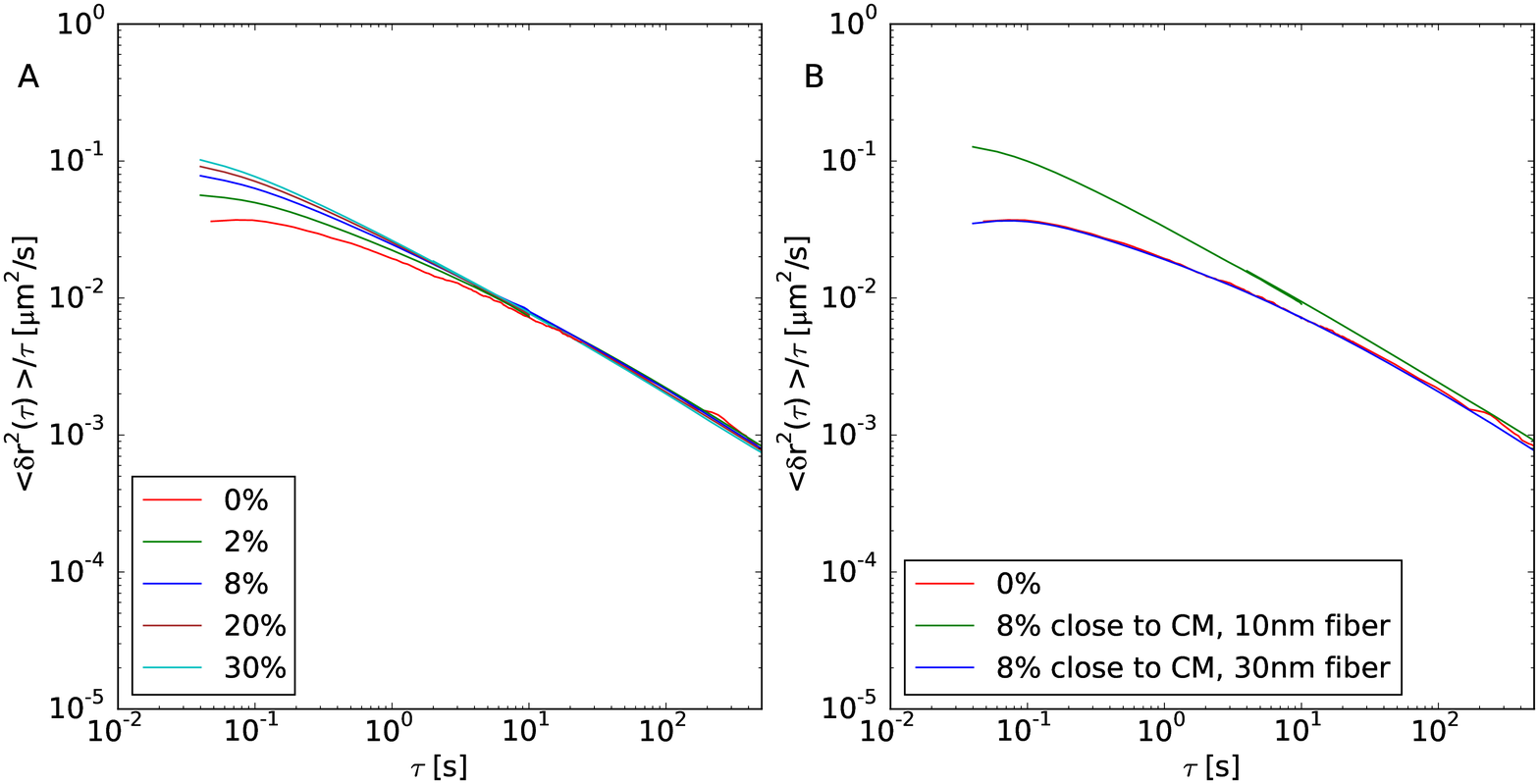}
\caption{{\bf Monomer mean-square displacement normalized by lag time $\tau$ $\langle \delta r^2(\tau) \rangle / \tau$.}
(A)
Case studies of model chromosomes made of 30nm fiber with randomly interspersed fragments of 10nm fiber.
The total amount of the 10nm fiber is indicated in the corresponding legend.
$\langle \delta r^2(\tau) \rangle$ has been averaged uniformly along the chain.
(B)
Cases studies of model chromosomes made of two separate domains of 10nm fiber ($8\%$ of the total chromosome size and positioned close to its centre of mass) and 30nm fiber,
with corresponding mean-squared displacements averaged separately on the two domains.
For comparison, we show in both panels the case of a chromosome made of a homogeneous 30nm fiber filament (red lines).
Mapping to real time ``$\tau_{MD}=0.02$ seconds'' follows from the discussion presented in reference~\cite{RosaPLOS2008}.}
\label{fig:LociMSD}
\end{figure}

Figure~\ref{fig:LociMSD} summarizes our results for $\langle \delta r^2(\tau) \rangle / \tau$ {\it vs.} $\tau$
for the two cases of random positioning of small filaments of 10nm fiber within the chromatin fiber (panel A) and for chromosomes made of two large separated domains with different fiber composition (panel B).
In both cases, we notice a general increase of chromatin mobility as larger and larger portions of 30nm fiber unfold,
and, at larger times, a trend which does not substantially depend on the small scale details of chromatin fiber.
Not surprisingly, data at short times reflect in part the discussed results for chromatin structure:
in particular, we notice that differences in chromatin mobility before and after chromatin unfolding in random locations are only visible below time-scales of about 5 seconds (panel A).
Slightly larger discrepancies are observed in the other situation where chromosomes are organized as two separate domains (panel B).
In this latter case, unfolded chromatin loci move on average more than folded ones, the latter displaying the same motion than in the case of the homogeneously folded chromosome (compare green {\it vs.} blue lines).

\paragraph*{Model verification: Experimental observation of local chromatin unfolding.}


In a recent work~\cite{TherizolsBickmoreScience2014}, Therizols {\it et al.} used FISH to show that, in embryonic stem cells, chromosome decondensation is sufficient to alter nuclear organization.
Two sets of experiments were done: first, a viral transactivator was used to activate transcription of three genes ($Ptn1$, $Nrp1$ and $Sox6$).
Alternatively, repositioning of the same genes was also observed after the treatment with an artificial peptide (DELQPASIDP) which decondenses chromatin without inducing transcription.
By comparing the measured spatial distances between the two ends of each selected sequence,
the authors concluded that nuclear organization is driven mainly by chromatin remodeling rather than transcription.

We have tested the predictions of our model by using data from these experiments.
We have simulated the unfolding of a chromatin region of genomic length corresponding to the size of the specific gene we wanted to mimic.
This situation corresponds to the second case studied in this work, namely the unfolding of a chromatin cluster.
For a fair comparison, we have processed the experimental data as follows:
we have  reconstructed first the probability density distribution function for spatial distances between the ends of each gene,
for the control condition (denoted as $eGFP$) and cells in which DELQPASIDP is recruited to the chromosome loci (denoted as $DEL$).
Since FISH distances are recorded as two-dimensional vectors projected on the confocal plane,
for the purpose of comparison a (large) set of three-dimensional distances with equivalent $2d$ projections
was generated numerically by assuming random orientations of the $3d$ vectors in relation to the axis orthogonal to the confocal plane.

It can be observed in figure \ref{fig:TherizolsData} panels A, C and E that, upon chromatin decondensation,
the peak {\it and} the shape of the distribution change dramatically, in particular the distributions become wider.
The same effect is displayed by our simulations (panels B, D and F).
This comparison thus validates our result that larger genes tend to expand when unfolded.

\begin{figure}[h]
\includegraphics[width=\textwidth]{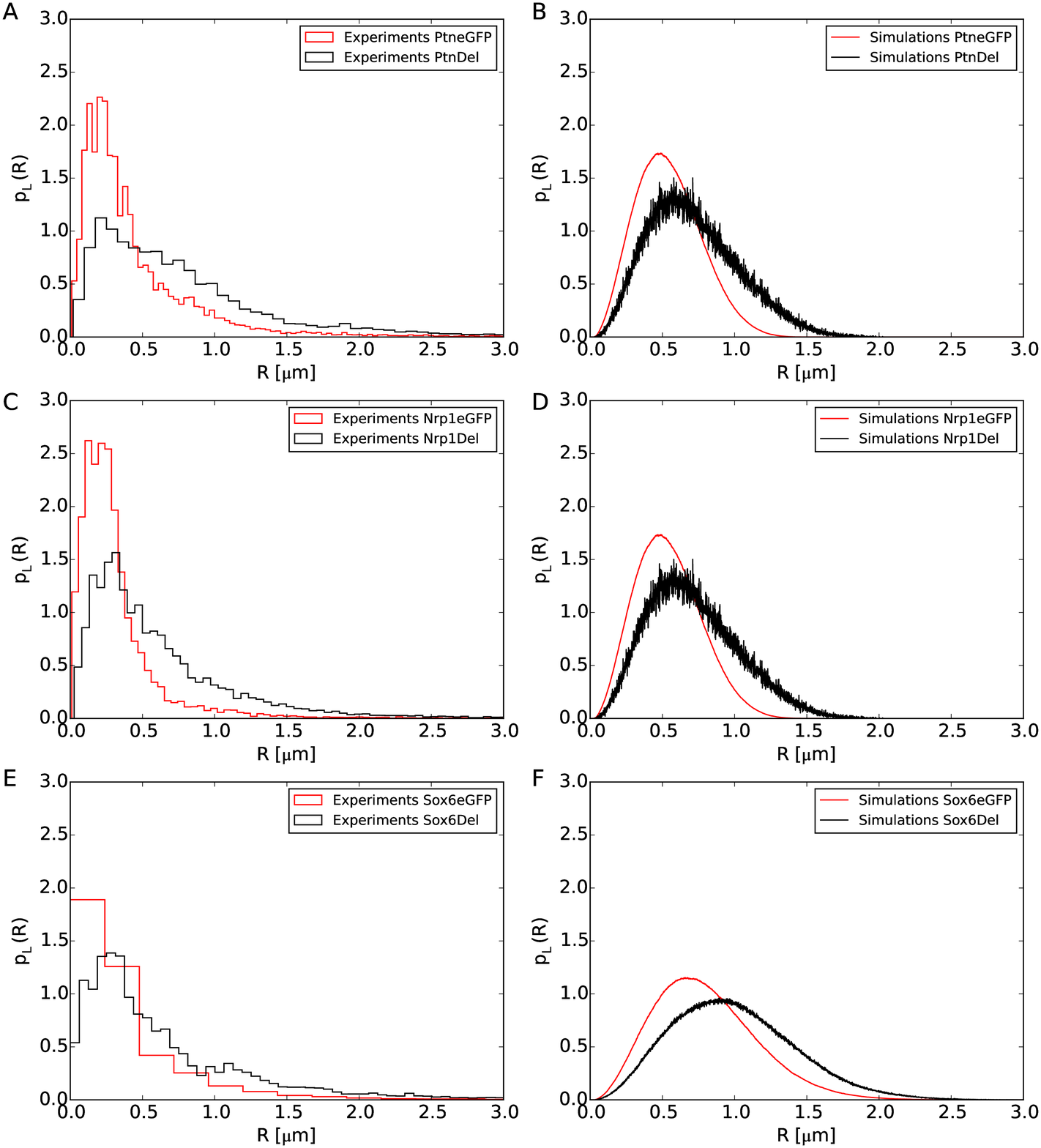}
\caption{{\bf Comparison of experimental~\cite{TherizolsBickmoreScience2014} (left) and simulation (right) data for the distribution functions, $p_{L} (R)$,
of spatial distances $R$ between the ends of the murine Ptn, Nrp1 and Sox6 genes.}
$L\sim 0.22, 0.21, 0.55$ Mbp is the genomic separation between the FISH probes placed at  the end of the three considered genes, respectively.
Experimental distances are measured before (eGFP, red) and after (DEL, black) stimulated chromatin decondensation.
Numerical results come from three independent molecular dynamics runs simulating the unfolding of chromatin filaments of corresponding sizes.
The rest of the chromosome remains in the 30nm fiber state.
}
\label{fig:TherizolsData}
\end{figure}

While our model describes reasonably well the trend of the experimental data,
the median values and the part of the distribution covering the larger distances, it seems to perform worse for the part of the distribution describing the small distances.
This could be due to several factors which, to maintain the model simple, have been neglected.
These include either protein linkers that physically bridge two sites along the genes forming a loop~\cite{NicodemiPombo2012,SanbornPNAS2015,MirnyLoopExtrusion2015},
either chemical modifications which could change the charge on some histones,
thus inducing an effective electrostatic attraction~\cite{JostVaillantNAR2014} between regions of the gene that would make it more compact.
Moreover, for the experiments using DEL, not all cells receive the exact same quantity of plasmid.
That is, the experimental distributions of distances could be biased towards small $R$ values because, in some cells, there was not enough plasmid to induce decondensation.

\paragraph*{Polymer physics aspects.}
The choice for the physical parameters employed in this work as fibers flexibilities, thicknesses and excluded volume interactions (see section ``Materials and Methods'') was mainly motivated by reasons of simplicity.
Understanding to what extent our results depend on these specific choices requires a deeper analysis.
In order to isolate and quantify the effects of one single parameter,
we have considered a simpler model chromosome where monomers have fixed nearest neighbor distance along the sequence equal to $30$ nm.
The polymer was then split in two complementary domains:
the fiber in one domain has the same features of the 30nm fiber, while the physical properties of the fiber in the other domain are changed, in turns, to:
(1) completely flexible fiber;
(2) thickness equal to 10 nm;
(3) larger strength for the corresponding excluded volume interaction.
Each of these changes was applied, in turns, to a domain occupying $20 \%$, $50 \%$ and respectively $100 \%$ of the polymer.
Figure 12 (top panels) shows that, not surprisingly, the persistence length influences $\langle R^2(L) \rangle$ and $\langle p_c(L) \rangle$ only at small scales up to $0.1$ Mbp,
suggesting that one of the main effects of having a heterogeneous chromatin composition consists in a global change of the overall persistence length of the chromatin fiber.
By varying instead the range of the excluded volume interactions without modifying the nominal persistence length we produce variations in $\langle R^2(L) \rangle$ and $\langle p_c(L) \rangle$
similar to the ones reported in figures~\ref{fig:inside} and~\ref{fig:outside} (middle panels).
Viceversa, when varying the strength of the interaction no change is observed (bottom panels).
These observations suggest a leading role for the effective range of the excluded volume interaction:
in our model, the thinner 10nm fiber tends to occupy a larger region because its own self-repulsion becomes less important when compared to the repulsion from the thicker 30nm fiber.
Quite interestingly, a similar effect concerning the change in the excluded volume was observed in equilibrium simulations of generic block copolymers~\cite{HansenBlockCopol2005}.
Finally, we stress that this effect is less visible in figure~\ref{fig:rsqcprob} because of the base pair content assigned to the monomers:
since the base pair content of the 10nm fiber region is in general low compared to 30nm fiber region, its effect is not visible when averaging over the entire chromosome.

\section*{Discussion}

In this article,
we have presented results of extensive Molecular Dynamics computer simulations of a coarse-grained polymer model for interphase chromosomes that extends previous numerical work~\cite{RosaPLOS2008,RosaBJ2010,DiStefanoRosa2013,ValetRosa2014} by introducing a crucial ingredient which was neglected before:
namely, the presence of two kinds of chromatin fibers of different thickness and stiffness mutually interacting inside their own chromosome territory.
Starting from a chromosome configuration made of a single and homogeneous 30nm fiber like filament,
we have monitored chromosome spatial and temporal behaviors when this conformation is altered by the introduction of increasing amounts of 10nm-like fiber through the controlled unfolding of the thicker 30nm fiber.

The work shows that there exists detectable chromosome (re)organization for spatial scales smaller than $0.1$ Mbp (figure \ref{fig:rsqcprob}) and time-scales shorter than just a few seconds (figure~\ref{fig:LociMSD}).
Quite interestingly, these findings appear systematic and do not depend on the size of the chromosome region affected by the phenomenon of local unfolding or by its location along the chromosome.

Interestingly, these results tend to suggest that experimental methodologies like FISH or HiC might be of little or no help in distinguishing between fibers of different compaction,
unless they investigate genomic distances smaller than $0.1$ Mbp.
This prediction can be tested, for instance, by employing the recently developed oligonucleotide based FISH probes which seem to provide the necessary fine resolution~\cite{Beliveau2012, Beliveau2015}.
An important conclusion of our work is that, although our model uses parameter values that can be associated to the traditional ``10nm/30nm'' chromatin fiber paradigm~\cite{MaeshimaEltsov,ReviewChromosoma2015},
our results reflect a generic physical effect~\cite{HansenBlockCopol2005} which ought to be observable in more general systems of crumpled polymers constituted of fibers with different thickness and/or stiffness (see section ``Polymer physics aspects'' and figure 9).

Our simulation protocol compares qualitatively well (see figure \ref{fig:TherizolsData}) with experimental results on chromosome reorganization in mouse embryos treated with a synthetic transcription factor~\cite{TherizolsBickmoreScience2014} which produces selective activation of a specific gene.
In particular, we predict the observed shifting for distribution functions of spatial distances.
Of course, it is quite possible that other mechanisms may explain the experimental results equally well.
In fact, intentionally our model tends to neglect other important aspects of chromosome folding which have been highlighted recently by other authors:
sequence-specific attractive interactions~\cite{JostVaillantNAR2014},
protein linkers between chromatin fibers~\cite{NicodemiPombo2012},
mechanisms of active regulation~\cite{BruinsmaGrosbergRabin2014,GanaiMenon2015},
``loop extrusion''~\cite{SanbornPNAS2015,MirnyLoopExtrusion2015} involved in the reorganization of small chromosome domains,
or the anchoring to the nuclear envelope or other nuclear organelles~\cite{RosaGehlen2006,GehlenLangowski2012,WongZimmer2012}.
In the lack of a more quantitative analysis, we can only speculate on the fact that the inclusion of these mechanisms into our model might alter significantly the conclusions sketched here.
On the other hand, this should represent an important stimulus for investigating further and more quantitatively the delicate relationship between chromosome structure and function.

\section*{Materials and Methods}

\subsection*{Simulation protocol I. Force field}


In this work, the chromatin fiber is modeled as a coarse-grained polymer chain,
with monomer-monomer interactions described by analytical expressions similar to the ones used in our previous works~\cite{RosaPLOS2008,RosaBJ2010,DiStefanoRosa2013}
and suitably adapted to take into account the different sizes of 30nm and 10nm monomers.

The full Hamiltonian governing the system, ${\cal H}$, consists of three terms:
\begin{eqnarray}\label{eq:IntraChainEnergy}
{\cal H} & = & \sum_{i=1}^N [ U_{FENE}(i, i+1) + U_{br}(i, i+1, i+2) + \sum_{j=i+1}^N U_{LJ}(i,j) ]
\end{eqnarray}
where $N$ (see Table~\ref{tab:RingSizeDecond}) is the total number of monomers constituting the ring polymer modeling the chromosome
(see Section ``Construction of model chromosome conformation'' for details on this point) and $i$ and
$j$ run over the indexes of the monomers. The latter are assumed to be
numbered consecutively along the ring from one chosen reference monomer.
The modulo-$N$ indexing is implicitly assumed because of the ring periodicity.

\begin{table}[!ht]
\caption{{\bf Summary of sizes of simulated model chromosomes.}
In our model, each chromosome is a mixture of 10nm and 30nm model chromatin fibers.
Column 1:
The total amount of 10nm chromatin fiber given as a fraction of the total chromosome size $=117.46 \times 10^6$ bp.
Column 2:
The total number of 30nm monomers ($N_{30}$) and 10nm monomers ($N_{10}$) of the model chromosome.
The total number of monomers per chromosome is $N=N_{30}+N_{10}$.
Column 3: The total length of the 30nm fiber portion and the 10nm fiber portion, in Mbp.
Column 4: The total length of the 30nm fiber portion and the 10nm fiber portion, in $\mu$m.}

\begin{tabular}{|c|c|c|c|}
\hline
\footnotesize{$\frac{\mbox{Total amount of 10nm fiber [bp]}}{\mbox{Chromosome size [bp]}}$} & \footnotesize{$N_{30} \, / \, N_{10}$} & \footnotesize{$L_{30} [\mbox{Mbp}] \, / \, L_{10} [\mbox{Mbp}]$} & \footnotesize{$L_{30} [\mu m] \, / \, L_{10} [\mu m]$}\\
\hline
0\% & $39154 \, / \, 0$ &$117.462 \, / \, 0$ & $1174.62 \, / \, 0 $\\
\hline
2\% & $38371 \, / \, 21141$ & $115.113 \, / \, 2.349$ & $1151.13 \, / \, 211.4 $\\

\hline
8\% & $36022 \, / \, 84564$ &  $108.066 \, / \, 9.396$ &$1080.66 \, / \, 845.64 $\\

\hline
20\% & $31323 \, / \, 211437$ & $93.969 \, / \, 23.493$ & $939.69 \, / \, 2114.37 $\\

\hline
30\% & $27678 \, / \, 316872$ & $83.034\, / \, 34.428$ & $830.34 \, / \, 3168.72 $ \\

\hline
100\% & $0 \, / \, 1057158$ & $0 \, / \, 117.462$ &$0 \, / \, 10571.58 $\\

\hline

\end{tabular}
\label{tab:RingSizeDecond}
\end{table}

By taking the nominal monomer diameter of the 30nm chromatin fiber, $\sigma = 30 \mbox{ nm} = 3000 \mbox{ bp}$~\cite{RosaBJ2010}, as our unit of length,
the vector position of the $i$th monomer, $\vec{r}_i$,
the pairwise vector distance between monomers $i$ and $j$, $\vec{d}_{i,j} = \vec{r}_j - \vec{r}_i$,
and its norm, $d_{i,j}$,
the energy terms in equation~\ref{eq:IntraChainEnergy} are given by the following expressions:

\noindent
1) The chain connectivity term, $U_{FENE}(i,i+1)$ is expressed as:
\begin{equation}\label{eq:fenepot}
U_{FENE}(i,i+1) = \left\{
\begin{array}{l}
- {k \over 2} \, R^2_0 \, \ln \left[ 1 - \left( {{d_{i,i+1}+\Delta} \over R_0} \right)^2 \right], \, d_{i,i+1} \leq R_0 - \Delta \\
0, \, d_{i,i+1} > R_0 - \Delta
\end{array}
\right.
\end{equation}
where $R_0=1.5 \sigma$, $k=30.0 \epsilon / \sigma^2$ and
the thermal energy $k_B\, T$ equals $1.0 \epsilon$~\cite{KremerGrestJCP1990}.
The shift distance $\Delta$ is used to tune the length of the bond between nearest neighbors monomers:
(a) $\Delta = 0$ if both beads model the 30nm fiber,
(b) $\Delta = \sigma / 3 = 10$ nm if one bead models the 30nm fiber and the other the 10nm fiber,
(c) $\Delta = 2 \sigma / 3 = 20$ nm if both beads model the 10nm fiber.

\noindent
2) The bending energy term has the standard Kratky-Porod form (discretized worm-like chain):
\begin{equation}\label{eq:stiffpot}
U_{br}(i, i+1 ,i+2) = \frac{k_B \, T \, \xi_p}{\sigma}  \left (1 - \frac{{\vec d}_{i,i+1} \cdot  {\vec d}_{i+1,i+2}}{d_{i,i+1}
 \, d_{i+1,i+2}} \right )
\end{equation}
The persistence length between triplets of 30nm beads is the same as in our previous works~\cite{RosaPLOS2008,RosaBJ2010,DiStefanoRosa2013}, $\xi_p = 5.0 \sigma = 150 \mbox{ nm}$.
In the other cases, we model the fiber as completely flexible, $\xi_p=0$.
This is in agreement with measurements showing that the persistence length for unfolded chromosomes is close to $\approx 6$ nm~\cite{bystricky}.

\noindent
3) The excluded volume interaction between distinct monomers (including consecutive ones)
corresponds to a purely repulsive Lennard-Jones potential:
\begin{equation}\label{eq:ljpot}
U_{LJ}(i,j) = \left\{
\begin{array}{l}
4 \epsilon \left[ \left(\frac{\sigma}{d_{i,j} + \Delta }\right)^{12} - \left( \frac{\sigma}{d_{i,j} + \Delta}\right)^6 + \frac{1}{4} \right], d_{i,j} \leq \sigma 2^{1/6} - \Delta\\
\\
0, d_{ij} > \sigma 2^{1/6} - \Delta
\end{array}
\right. ,
\end{equation}
where the shift distances $\Delta$ take the same values as the ones adopted in the $U_{FENE}$ term described above.

\subsection*{Simulation protocol II. Molecular Dynamics simulations}

As in our previous work~\cite{RosaPLOS2008,RosaBJ2010,DiStefanoRosa2013,ValetRosa2014},
chromosome dynamics was studied by performing fixed-volume Molecular Dynamics (MD) simulations with periodic boundary conditions
at near-physiological fixed chromatin density $\rho = 0.012 \mbox{ bp} / \mbox{nm}^3$.

Note that periodic boundary conditions {\it do not} introduce confinement to the simulation box:
using properly unfolded coordinates, the model chromatin fibers can extend over arbitrarily large distances~\cite{RosaPLOS2008}.

The system dynamics was integrated by using LAMMPS~\cite{lammps}
with Langevin thermostat in order to keep the temperature of the system fixed to $1.0 k_B T$.
Given the unit mass $m_{30} = 1$ of the 30nm-bead, we fixed the mass of the 10nm-bead to $m_{10} = \frac{1}{27}$.

The integration time step was fixed to $t_{int} = 0.001 \tau_{MD}$,
where $\tau_{MD}=\sigma \left( \frac{m_{30}}{\epsilon} \right)^{1/2}$ is the elementary Lennard-Jones time.
$\gamma = 0.5 / \tau_{MD}$ is the friction coefficient~\cite{KremerGrestJCP1990}
which takes into account the corresponding interaction with the background implicit solvent.
The total length of each MD simulation run is $=3 \cdot 10^5 \, \tau_{MD}$,
with an overall computational effort ranging from a minimum of $\approx 10^3$ to a maximum of $\approx 9 \cdot 10^4$ hours of single CPU for ``$0\%$'' and ``$100\%$'' model chromosome conformations, respectively.
Single chromosome conformations were sampled every $10^3 \tau_{MD}$, implying $300$ configurations per each run.

We have verified that our rings are well equilibrated by considering the mean-square distances, $\langle R^2(L) \rangle$, and mean contact probabilities, $\langle p_c(L) \rangle$,
calculated on different sets of configurations log-spaced in time.
As shown in Figure 13 for a $20\%$ amount of 10nm fiber, all curves give the same results.
Similar results are obtained for all simulated systems.

\subsection*{Simulation protocol III. Initial configuration}

\paragraph*{Construction of the model chromosome conformation.}

In spite of the complexity of the chromatin fiber and the nuclear medium,
three-dimensional chromosome conformations are remarkably well described by generic polymer models~\cite{Grosberg_PolSciC_2012,GrosbergKremerRev_RPP2014,RosaZimmer2014}.
In particular, it was suggested~\cite{RosaPLOS2008,VettorelPhysToday2009} that the experimentally observed~\cite{hic} crumpled chromosome
structure can be understood as the consequence of slow equilibration of chromatin fibers due to mutual chain uncrossability during thermal motion.
As a consequence, chromosomes do not behave like equilibrated {\it linear} polymers in solution~\cite{DoiEdwards,RubinsteinColby}.
Instead, they appear rather similar to unlinked and unknotted circular (ring) polymers in entangled solution.
In fact, under these conditions ring polymers are known to spontaneously segregate and
form compact conformations~\cite{VettorelPhysToday2009,GrosbergKremerRev_RPP2014,RosaEveraersPRL2014},
strikingly similar to images of chromosomes in live cells obtained by fluorescence techniques~\cite{CremerReview2001}.

Due to the typically large size of mammalian chromosomes (each chromosome contains, on average, $\sim 10^8$ basepairs of DNA),
even minimalistic computational models would require the simulation of large polymer chains,
made of tens of thousands of beads ~\cite{RosaPLOS2008,RosaBJ2010,DiStefanoRosa2013}.
For these reasons, in this work we resort to our recent mixed Monte-Carlo/Molecular Dynamics multi-scale algorithm~\cite{RosaEveraersPRL2014}
in order to design a {\it single, equilibrated} ring polymer conformation at the nominal chromatin density of $\rho = 0.012 \mbox{ bp} / \mbox{nm}^3$ (see previous section).
The ring is constituted by $N=39154$ monomer particles of linear size $\sigma = 30$ nm $=3000$ basepairs,
which corresponds to the average linear size of a typical mammalian chromosome of $\approx 117.46 \times 10^6$ basepairs.
By construction, the adopted protocol guarantees that the polymer has the nominal local features of the 30nm-chromatin fiber
(stiffness, density and topology conservation) that have been already employed elsewhere~\cite{RosaPLOS2008,RosaBJ2010,DiStefanoRosa2013,ValetRosa2014}.
For the details of the multi-scale protocol, we refer the reader to reference~\cite{RosaEveraersPRL2014}.

\paragraph*{Local unfolding of chromatin fiber.}

\begin{figure}[h]
\includegraphics[width=\textwidth]{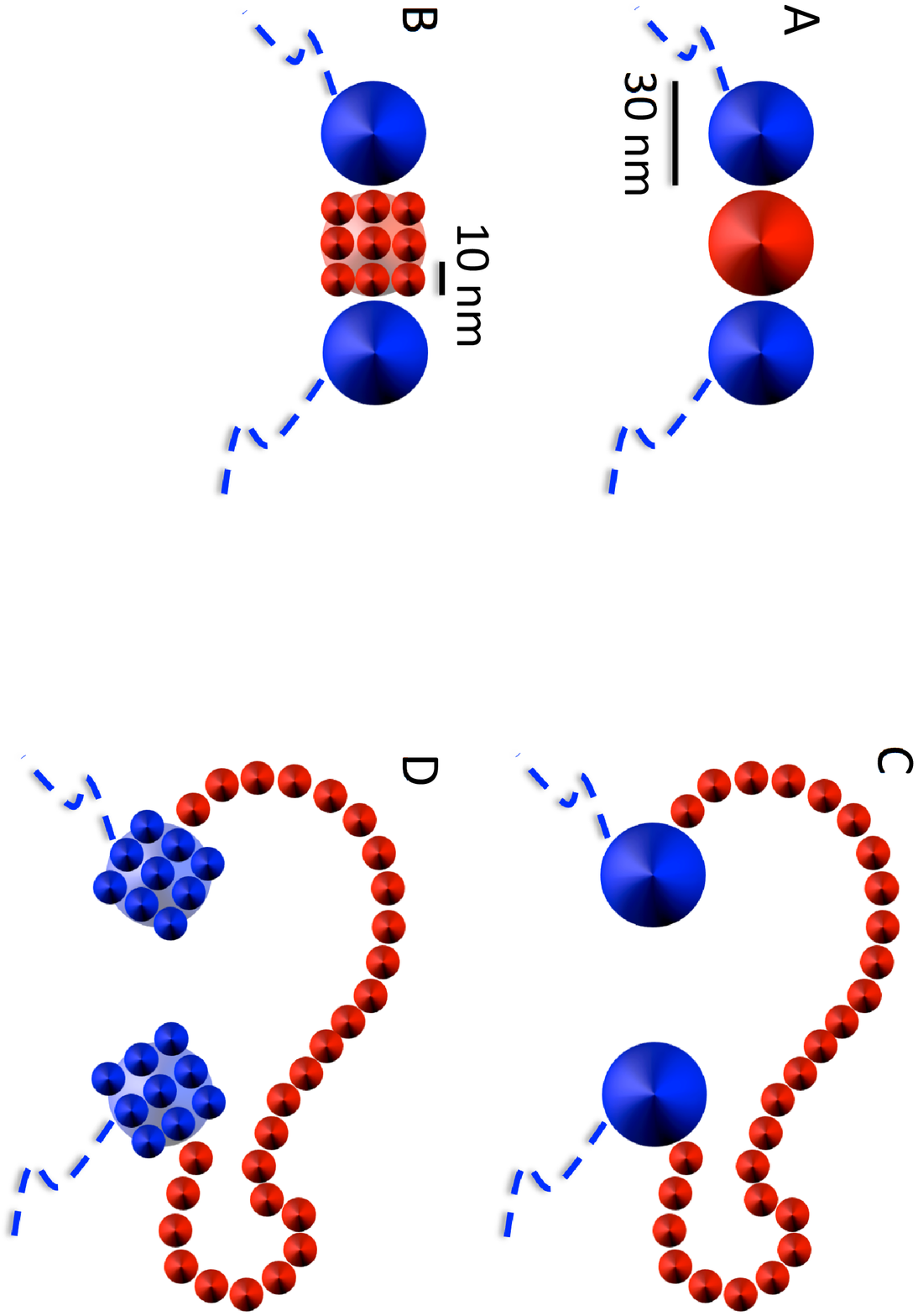}
\caption{{\bf Illustration of the adopted set-up used to model the unfolding of the 30nm model chromatin fiber into the 10nm model chromatin fiber.} 
Each bead of 30 nm linear size selected for the unfolding procedure (A, in red for clarity) is substituted (B) by 27 smaller beads of linear size $=10$ nm arranged on a $3\times3\times3$ regular cubic lattice (the figure shows only one face, for simplicity).
(C)
During swelling, the small monomers move away from the initial cubic conformation and assume a random-walk like conformation.
(D)
Average quantities which depend on genomic separations $L$, like $\langle R^2(L) \rangle$ and $\langle p_c(L) \rangle$,
are calculated by measuring $L$ in 10nm fiber units after replacing all 30nm fiber monomers not involved in the MD decondensation step by $27$ equivalent 10nm monomers.
}
\label{fig:Unfolding}
\end{figure}

To model the 30nm fiber, we employ the typical density of $\approx 100$ bp/nm~\cite{bystricky}, which we have already adopted successfully elsewhere~\cite{RosaPLOS2008,RosaBJ2010,DiStefanoRosa2013,ValetRosa2014}.
For the 10nm fiber,
we assume that the typical repeat length of a nucleosome ($\approx 200$ bp)
is shared between the nucleosome core particle ($147$ bp of DNA for $10$ nm of diameter)
and linker DNA ($\approx 60$ bp) assumed completely stretched ($\approx 20$ nm).
The total compaction of the 10nm fiber results being then $\approx 200 \mbox{ bp} \, / \, (10 \mbox{ nm} + 20 \mbox{ nm}) \approx 7$ bp/nm.

As shown next, and for practical purposes only, we round this quantity to $\approx 11$ bp/nm implying that each 30nm-fiber monomer (=$3000$ bp) swells into 27 10nm-fiber monomers.

In fact, in order to model the effect of having alternate filaments of 10nm and 30nm fibers within the same chromosome,
we started from the homogeneous model chromosome constructed as outlined in the previous section
and we gradually replaced each single 30nm fiber monomer by $3 \times 3 \times 3 = 27$ beads of smaller diameter $=10$ nm arranged on a regular cubic lattice of linear step $=10$ nm,
as shown schematically in figure~\ref{fig:Unfolding} (panels A and B).
This step was followed by a short ($120~\tau_{MD}$) MD run employing the LAMMPS option NVE/LIMIT (http://lammps.sandia.gov)
which, by limiting the maximum distance a particle can move in a single time-step,
allows the accommodation of the new beads without introducing mechanical stress inside the structure.
Next, we performed one more short run of $120~\tau_{MD}$ with the normal thermostat to accommodate the monomers further.
The resulting conformation was then employed as the input configuration for the full MD production run.

It is important to stress that the number of $10$nm fiber monomers and their content in base pairs are not equivalent (see Table~\ref{tab:RingSizeDecond}).
In fact, the unfolding of a given quantity of $30$nm fiber monomers produces a comparable much longer (in $\mu$m) $10$nm fiber:
upon unfolding and equilibration, the $3000$ basepair content lumped in a $30$nm monomer will now extend over a total contour length of $270$ nm.
The total length (in $\mu$m) of the chromosome changes upon decondensation, and a total amount of $2 \%$ of 10nm fiber only will now extend over $\sim 18\%$ of the total contour length.
For the largest content of $30 \%$ of 10nm fiber studied in this work, the corresponding ratio to the total chromosome contour length is $\sim 80 \%$.
Given the fact that, from the physical point of view, is the relative contour length of the blocks rather than the relative base pair content that influences the behaviour,
we assume having covered a sufficient range of case studies.

\begin{figure}[h]
\includegraphics[width=\textwidth]{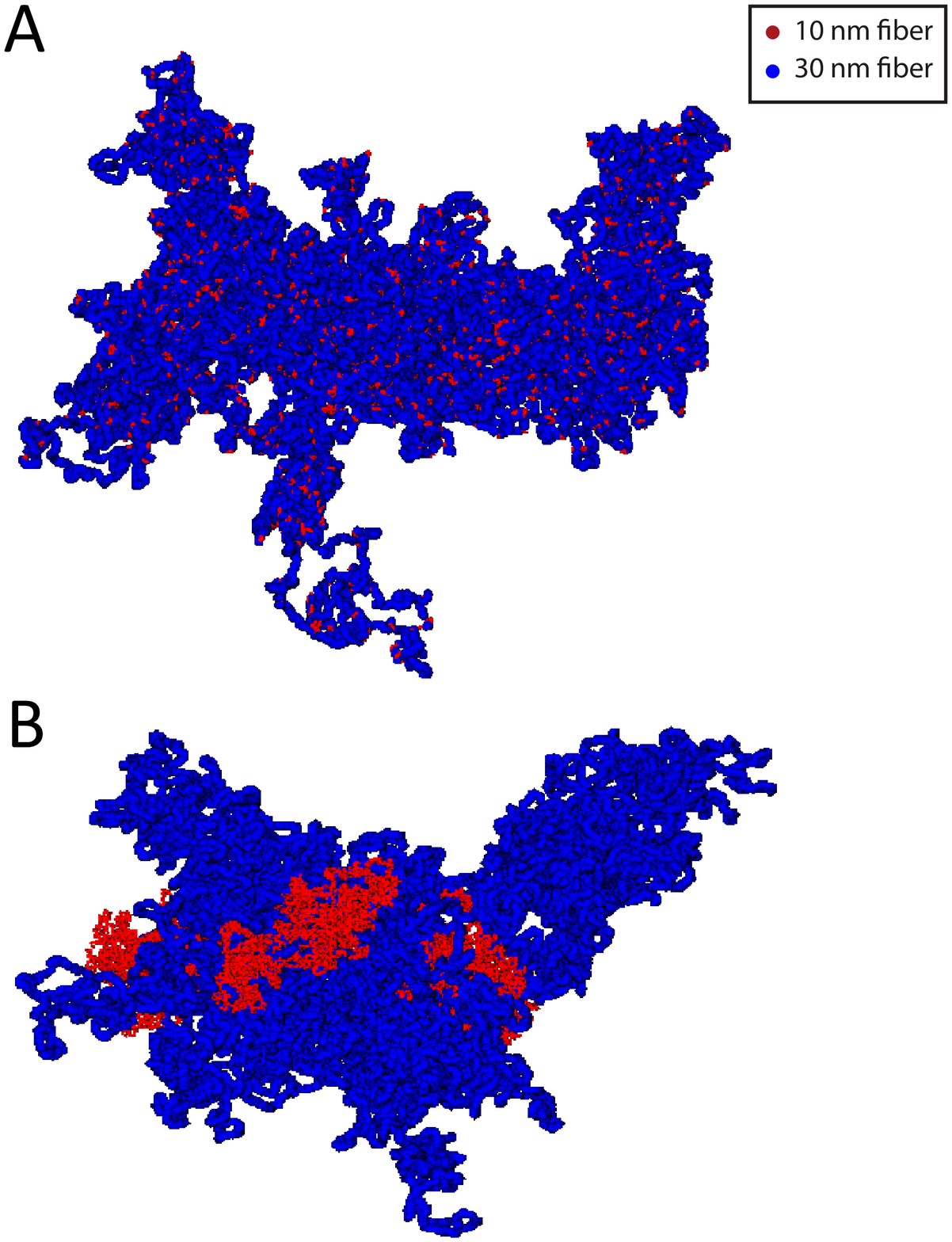}
\caption{{\bf Model chromosome conformations with a $8\%$ total amount of 10nm chromatin fiber.} 
(A)
Model conformation containing short, randomly disperse 10nm fiber sequences.
(B)
Model conformation containing a continuous sequence of 10nm fiber localized close to the chromosome centre of mass.}
\label{fig:ChrSnaps}

\end{figure}

In this work, we have studied the effects of the local unfolding of chromatin fiber in two different scenarios:
\begin{enumerate}
\item
In the first one, the 30nm fiber beads selected for being decondensed into 10nm fiber beads were chosen at random positions along the chromatin fiber,

with corresponding amounts summarized in Table~\ref{tab:RingSizeDecond}.
Figure~\ref{fig:ChrSnaps}A shows a snapshot of a model chromosome conformation with $8\%$ of 10nm fiber content.
\item
In the second scenario, they were chosen clustered along a single linear chromatin sequence.
Here, two prototypical situations were considered:
either the sequence is centered around the 30nm monomer at the closest distance to the chromosome centre of mass,
or the sequence is centered around the monomer at the farthest distance from the center of mass.
In this way, we were able to monitor the two very distinct situations where chromatin unfolding is located inside the corresponding chromosome territory or at its periphery.

A summary of the simulated systems is given in Table~\ref{tab:RingSizeDecond}.
Figure~\ref{fig:ChrSnaps}B shows a snapshot of a model chromosome conformation with $8\%$ of 10nm fiber content localized close to the chromosome centre of mass.
\end{enumerate}
For reference, we have also considered the two limiting cases where all monomers are either 30nm fiber or 10nm fiber like (Table~\ref{tab:RingSizeDecond}).

\subsection*{Data analysis}
Quantities discussed in this work as the mean-square distances, $\langle R^2(L) \rangle$, and the average contact frequencies, $\langle p_c(L) \rangle$, between chromosome loci are plotted as a function of the genomic distance, $L$.
In order to avoid unphysical behavior arising from the ring closure condition,
we considered contour lengths $L \leq 1/4$ of the {\it total} contour length of the ring, or $L \leq 30$ Mbp.

Possible numerical artifacts due to the presence of monomers with different degrees of resolution (``10nm fiber'' monomers {\it vs.} ``30nm fiber'' monomers)
were removed by averaging over all possible pairs of monomers at fixed genomic separations $L$ and spatial resolution of the 10nm fiber:
this was achieved by replacing all 30nm fiber monomers not already decondensed by the $27$ equivalent 10nm monomers, as shown in figure~\ref{fig:Unfolding} (panels C and D). 
In this way, each chromosome conformation always contributes with the same number of monomers
and genomic distances smaller than 3 kbp can be effectively sampled.
It is important to stress that what is described here concerns only the final analysis of the data,
and it is not implicated in the motion of the monomers during the MD runs which is always performed as described in previous sections.

Final values for $\langle R^2(L) \rangle$ and $\langle p_c(L) \rangle$ at large $L$'s were obtained by averaging further over {\it log-spaced} intervals centered at the corresponding $L$'s.
This procedure improves the accuracy by reducing considerably statistical fluctuations.
Error bars were calculated accordingly.


\section*{Acknowledgments}
The authors are indebted to K. Bystricky and V. Zaburdaev for stimulating discussions and suggestions.
AMF and AR acknowledge computational resources from SISSA HPC facilities.

\nolinenumbers

%
%
%
\bibliographystyle{plos2015} 
\bibliography{biblio}



\begin{figure}[h]
\includegraphics[width=\textwidth]{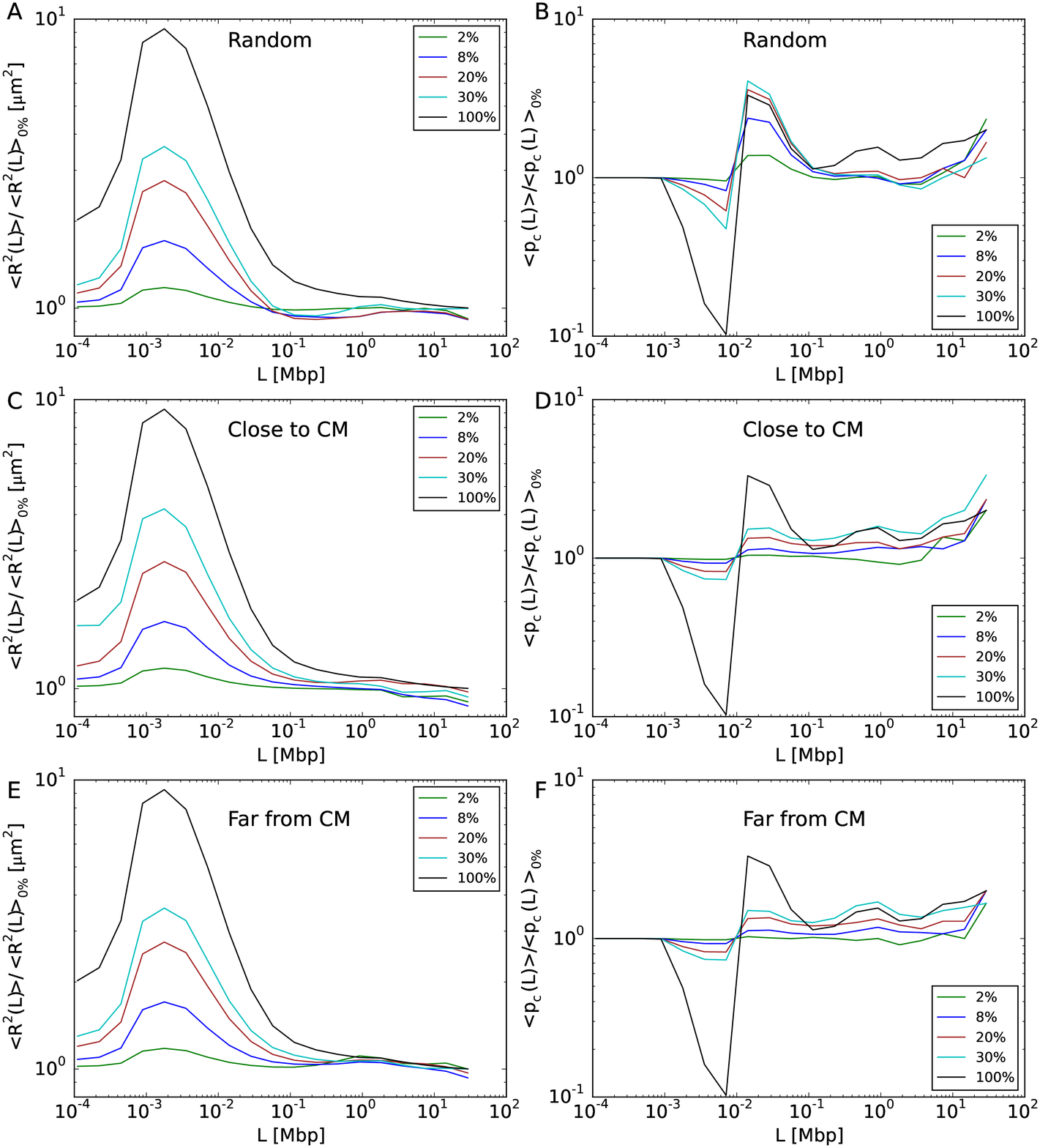}
\caption{{\bf Structural properties of model chromosomes composed of both 10nm fiber and 30nm fiber.}
Same data as in figure~\ref{fig:rsqcprob} of the main text, represented as ratios to corresponding mean-square distances and mean contact frequencies (left and right panels, respectively) of model chromosomes made of 30nm fiber.}
\label{fig:ShowDecondensation1}
\end{figure}

\begin{figure}[h]
\includegraphics[width=\textwidth]{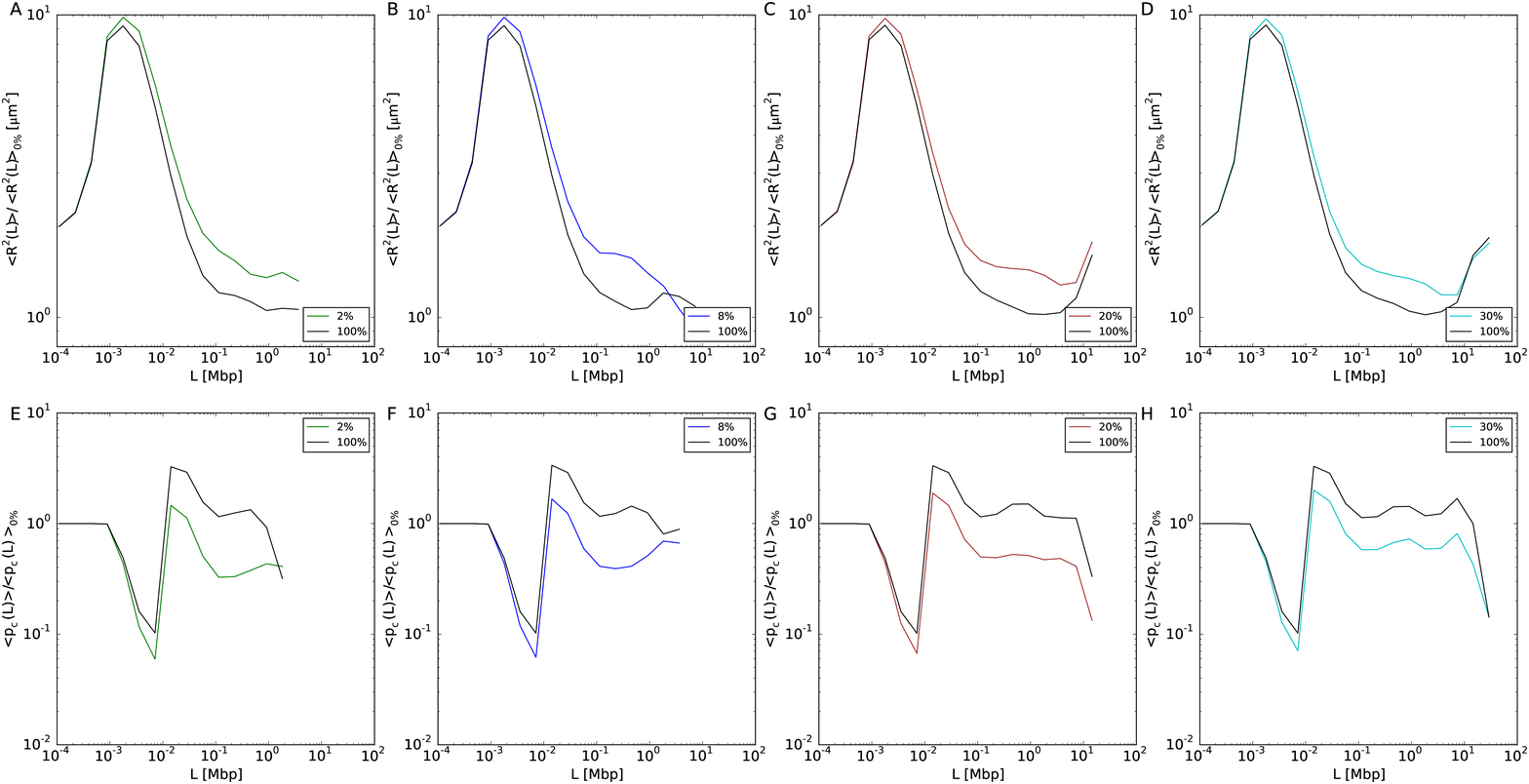}
\caption{{\bf Structural properties relative to the 10nm portion of model chromosomes composed of two separate domains of 10nm fiber and 30nm fiber, with the 10nm domain positioned closer to the chromosome center of mass.}
Same data as in figure~\ref{fig:inside} of the main text, represented as ratios to corresponding mean-square distances and mean contact frequencies (top and bottom panels, respectively) of model chromosomes made of 30nm fiber.}
\label{fig:ShowDecondensation2}
\end{figure}

\begin{figure}[h]
\includegraphics[width=\textwidth]{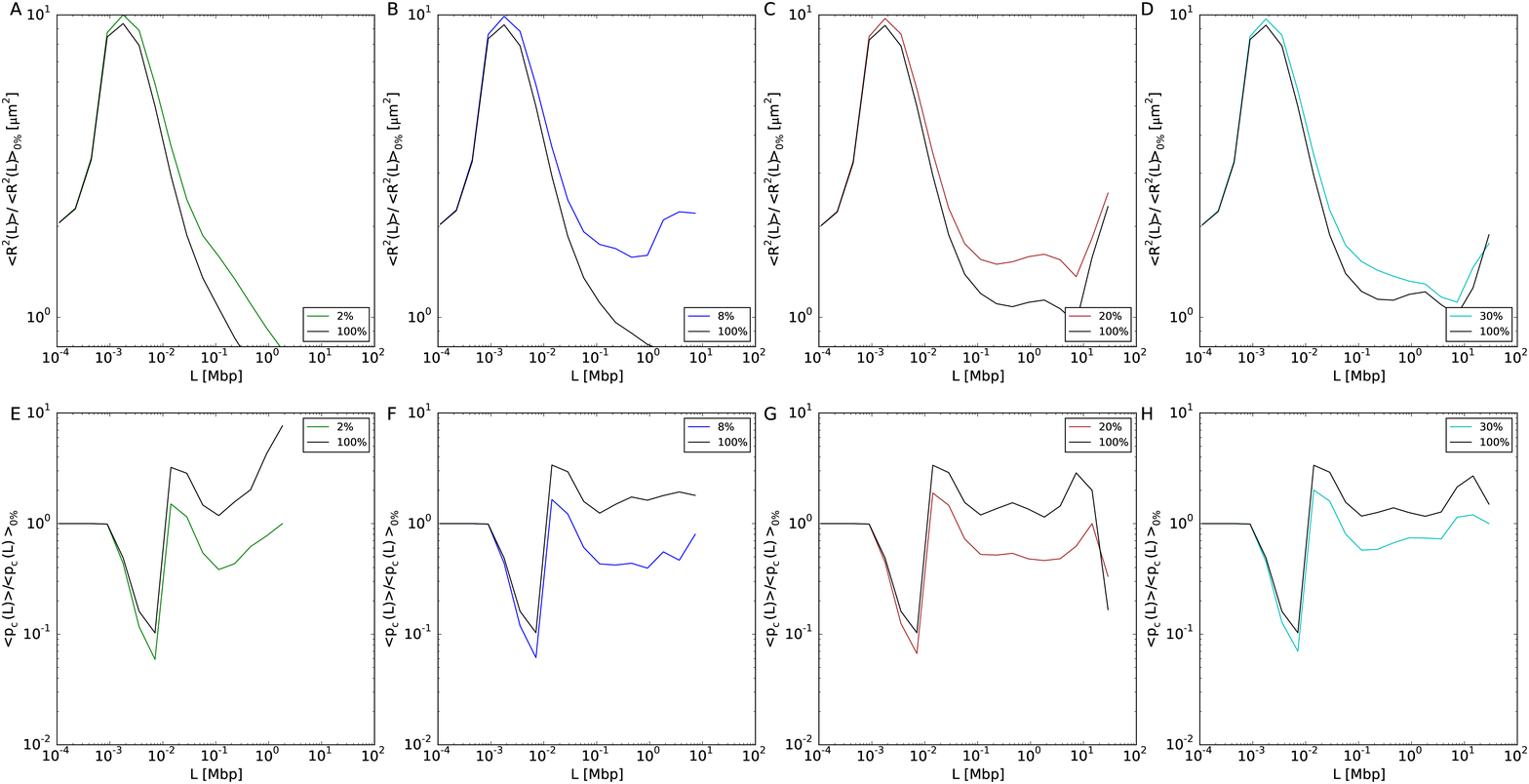}
\caption{{\bf Structural properties relative to the 10nm portion of model chromosomes composed of two separate domains of 10nm fiber and 30nm fiber, with the 10nm domain positioned farther from the chromosome center of mass.}
Same data as in figure~\ref{fig:outside} of the main text, represented as ratios to corresponding mean-square distances and mean contact frequencies (top and bottom panels, respectively) of model chromosomes entirely made of 30nm fiber.}
\label{fig:ShowDecondensation3}
\end{figure}

\begin{figure}[h]
\includegraphics[width=\textwidth]{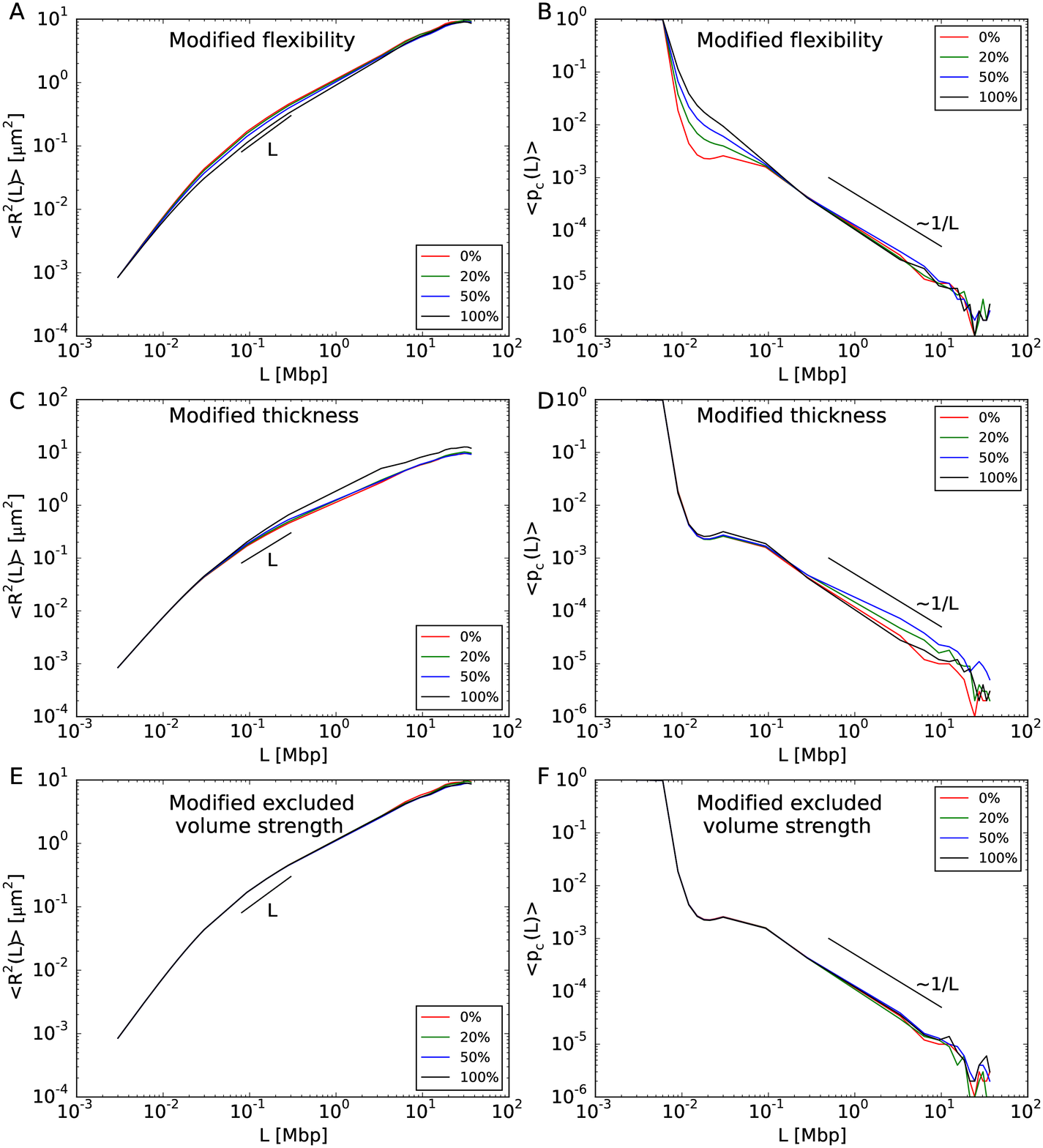}
\caption{{\bf Mean-square internal distances, $\langle R^2(L) \rangle$, and contact frequencies, $\langle p_c(L) \rangle$, between chromosome loci at genomic separation $L$.}
Results for model chromosome conformations where monomers have fixed nearest neighbor distance along the chromatin sequence equal to $30$ nm.
The polymer is split in two complementary domains of different sizes.
The fiber in one domain has the standard physical properties of the 30nm fiber.
For the other domain, we explore different linear sizes (shown in the caption in {\it percent} of the total chromosome size) and we modify in turns the following physical properties of the corresponding fiber:
(top) full flexibility;
(middle) reduced thickness $=10$ nm;
(bottom) increased strength of repulsive monomer-monomer interactions.}
\label{fig:polymer}
\end{figure}

\begin{figure}[h]
\includegraphics[width=\textwidth]{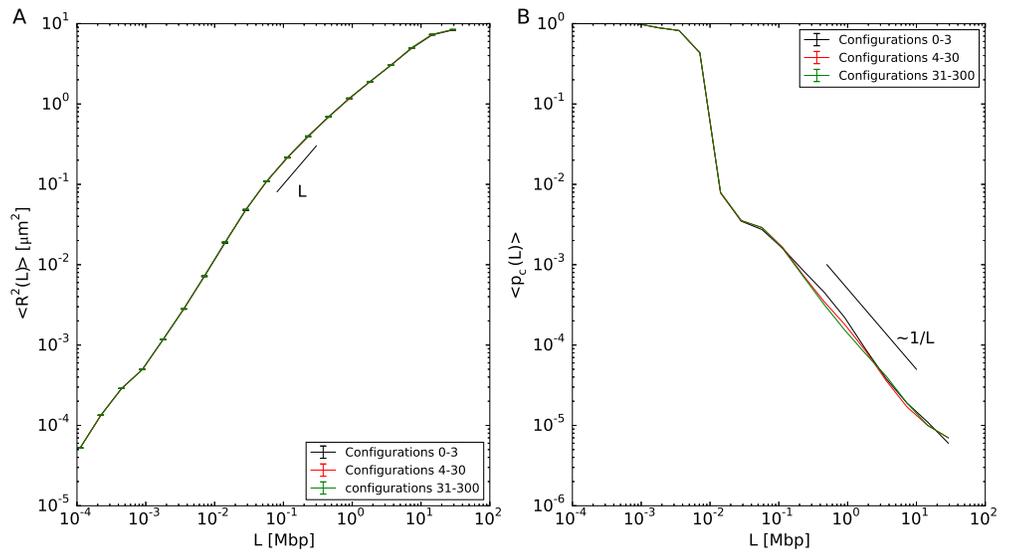}
\caption{{\bf Equilibration of chromosome conformations.}
Results for the mean-square internal distances, $\langle R^2(L) \rangle$, and mean contact frequencies, $\langle p_c(L) \rangle$, between chromosome loci at genomic separation $L$,
and a continuous filament of 10nm fiber occupying the $20\%$ of the chromosome.
The three curves of each plot show separate averages over log-spaced time intervals.}
\label{fig:equilibration}
\end{figure}

\end{document}